\documentclass[12pt,a4paper]{article}
\usepackage{graphicx}
\usepackage{verbatim}
\usepackage{longtable}
\usepackage{dcolumn}
\usepackage{harvard}
 \usepackage{setspace}

\newcommand{\req}[1]{(\ref{#1})}
\newcommand{\be}{\begin{equation}}
\newcommand{\ee}{\end{equation}}
\newcommand{\bea}{\begin{eqnarray}}
\newcommand{\eea}{\end{eqnarray}}

\newcommand{\pr}[1]{\left(#1\right)}
\newcommand{\cro}[1]{\left[#1\right]}

\newcommand{\avg}[1]{\langle{#1}\rangle}
\newcommand{\ovl}[1]{\overline{#1}}
\newcommand{\BE}{\begin{eqnarray}}
\newcommand{\EE}{\end{eqnarray}}
\newcommand{\BEn}{\begin{eqnarray*}}
\newcommand{\EEn}{\end{eqnarray*}}
\newcommand{\barr}{\begin{array}}
\newcommand{\earr}{\end{array}}

\newcommand{\bit}{\begin{itemize}}
\newcommand{\eit}{\end{itemize}}
\newcommand{\bc}{\begin{center}}
\newcommand{\ec}{\end{center}}
\newcommand{\ben}{\begin{enumerate}}
\newcommand{\een}{\end{enumerate}}

\newcommand{\eps}{\epsilon}

\newcommand{\sign}{\rm{sign}}

\begin{document}

\title{The tick-by-tick dynamical consistency of price impact in limit order books}
\author{Damien Challet\\
Physics Department\\University of Fribourg,  P\'erolles\\
1700 Fribourg, Switzerland\\
damien.challet@unifr.ch\\
Phone+Fax +41 26 300 91 \bigskip
\\
{\em previously at}\\
Institute for Scientific Interchange\\
Viale S. Severo 65\\
10133 Torino, Italy
}
\date{\today}

\maketitle

\begin{abstract}
Constant price impact functions, much used in financial literature, are shown to give rise to paradoxical outcomes since they do not allow for proper predictability removal: for instance the exploitation of a single large trade whose size and time of execution are known in advance to some insider leaves the arbitrage opportunity unchanged, which allows arbitrage exploitation multiple times. We argue that chain arbitrage exploitation should not exist, which provides an {\em a contrario} consistency criterion. Remarkably, all the stocks investigated in Paris Stock Exchange have dynamically consistent price impact functions. Both the bid-ask spread and the feedback of sequential same-side market orders onto both sides of the order book are essential to ensure consistency at the smallest time scale.

\end{abstract}

\section{Introduction}
There are several starting points for studying financial market dynamics
theoretically. The most common approach consists in assuming that the actions of
traders and external news result collectively into rather unpredictable price fluctuations. This view, dating back at least to
\citeasnoun{bachelier}, is the basis of the Efficient Market Hypothesis (EMH)
\cite{Fama}, which states more precisely that there is no risk-free arbitrage, thus, that it should be impossible to consistently outperform the market. In practice, for the sake of simplicity, it is usual to assume that the price evolution is completely unpredictable, i.e., to model it as a random walk. This is of course very convenient for elaborating option pricing theories \cite{HullBook}. Behavioural finance
starts from EMH and incorporates deviations from rational expectations in
the behaviour of the agents  in an attempt to explain anomalous
properties of financial markets (see \citeasnoun{BehavioralFinance} for a
recent review). Other approaches consist to find an empirical arbitrage opportunity and determine the optimal investment size \cite{ChenLimitArbitrage}, or to inject a controllable amount of predictability into a toy market model and study how the traders exploit it and remove it  in order to understand theoretically the dynamical
road to efficiency \cite{MMM,C05}.

Even the champions of EMH agree that there are temporary anomalies,
due for instance to uninformed traders (sometimes called noise traders),
 that arbitrageurs tend to cancel. Thus the
random walk hypothesis must be viewed as an extreme assumption
describing an average idealised behaviour that does not describe every
detail of the microscopic price dynamics. And indeed extreme assumptions are
most useful in any theoretical framework. This is why the opposite
assumption is worth considering: the discussion will start by focusing
on a single transaction and investigates how to exploit perfect
knowledge about it. Trader $0$ is active at time $t$; he buys/sells a
given amount of shares $n_0$,  leading to (log-)price change $r(t)=r_0$, where
$t$ is in transaction time, $t$ being the $t$-th transaction. Trader $1$ is assumed to have  perfect information about $t$ and
$r_0$, and act accordingly.

A related situation is found in \citeasnoun{FarmerForce}, which discusses
 the exploitation of an isolated pattern; its main
result is that the pattern does not disappear; on the contrary, its very
exploitation spreads it or shift it around $t$. Therefore, this work raises the question of how microscopic arbitrage removal is possible at all. Even if such interrogation may appear as rather incongruous, what follows will make it clear that assuming the absence of arbitrage does not imply that one understand how market participants achieve it. Even more, since  the arbitrage provided by the isolated exploitable pattern still exists and has the same intensity after having been exploited, it can be exploited again and again; this strongly suggests the emergence of a paradox which questions the validity of usual assumptions about price impact.

This paper aims at answering these two questions. It shows first the
need to change the way price impact and speculation are incorporated into
financial literature. Indeed the latter very often relies on two
fundamental assumptions: constant and symmetric price impact
functions. The second point is how arbitrage is removed. Most of current literature
restricts its attention to single-time price returns, without taking
into account the fact that speculation is inherently inter-temporal. This is simpler but wrong. Indeed, one does not make money by transacting, but by holding a
position, that is, by doing mostly nothing.

Both assumptions are useful simplifications that made possible some
understanding of the dynamics of market models. However, since the
discussion on arbitrage in market models is generally in discrete time,
one could in principle argue that one time step is long enough to
include holding periods, but this is inconsistent with the nature of
most financial markets. Indeed, the buy/sell orders arrive usually
asynchronously in continuous time, which rules out the possibility of
synchronous trading; at a larger time scale, the presence of widely
different time scales of market participants also rules out perfect synchronicity.

Once arbitrage exploitation is considered at its most minute temporal
level, that is, when its ineluctable inter-temporality is respected, a
simple consistency criterion for price impact functions emerges. If
violated, a paradoxical arbitrage chain exploitation is possible, or equivalently, the gain of a money-making trader is not decreased if he informs his trusted friends of the existence of arbitrage opportunities.
A weaker consistency criterion for price impact functions with respect to price manipulation is found in \citeasnoun{HubermanStanzl}; the discussion, that includes the case of time-varying linear price impact functions, does not take into account  the feedback of market orders on the order book, and only indirectly the spread. As we shall see,
the presence of the spread and the dynamics of price impact functions are key elements of
consistency. A more recent and sophisticated approach studies how to an arbitrage-free condition links the shape and memory of market impact \cite{gatheral-no}.

By focusing on the exploitation a single transaction isolated in time,
and by assuming that trader $1$ experiences no difficulty in injecting
his transactions just before and after $t$, his risk profiles with
respect to price fluctuations and uncertain position holding period are mostly irrelevant and will be neglected.

\section{The paradox}

The price impact function $I(n)$ is by definition the relative price
change caused by a transaction of $n$ (integer) shares ($n>0$ for
buying, $n<0$ for selling); mathematically,

\be
p(t+1)=p(t) + I(n),
\ee

\noindent where $p(t)$ is the {\em log}-price and $t$ is in transaction
time. We shall first assume that there is no spread, i.e. that when a market order moves the price, the spread closes immediately; spread alone cannot solve the paradox, as explained later.

The above notations misleadingly suggests that $I$ does not depend
on time. In reality, not only $I$ is subject to random fluctuations
(which will be neglected here), but also, for instance, to reactions to market order sides, which have a long
memory (see e.g. \citeasnoun{BouchaudLimit3} \citeasnoun{FarmerSign} \citeasnoun{BouchaudLimit4} \citeasnoun{FarmerMolasses} for
discussions about the dynamical nature of market impact). Neglecting the
dynamics of $I$ requires us to consider specific shapes for $I$ that
enforce some properties of price impact for each transaction, whereas in
reality they only hold on average. For example, one should restrict
oneself to the class of functions that makes it impossible to obtain
round-trip positive gains \cite{FarmerForce}. But the inappropriateness
of constant price impact functions is all the more obvious as soon as
one considers how price predictability is removed by speculation, which
is inter-temporal by nature.

The most intuitive (but wrong) view of market inefficiency is to regard
price predictability as a scalar deviation from the unpredictable case:
if there were a relative price deviation $r_0$ caused by a transaction
of $n_0$ shares at some time $t$. According to this view, one should
exchange $n_1$ shares so as to cancel perfectly this anomaly, where
$n_1$ is such that  $I(n_1)=-r_0$. This view amounts to regarding
predictability as something that can be remedied with a single trade.
However, the people that would try and cancel $r_0$ would not gain
anything by doing it unless they are market makers who try to stabilise
the price. The speculators on the other hand make money by opening,
holding, and closing positions. Hence one needs to understand the
mechanisms of arbitrage removal by the speculators.

\begin{figure}
\centerline{\includegraphics[width=0.7\textwidth]{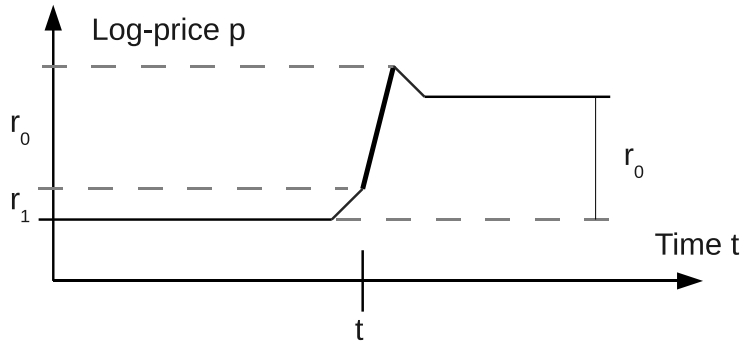}}
\caption{Price time dynamics around $t$ when trader 1 exploits his knowledge about $r_0$. This figure is readily generalised to the case of trader 1 and his friends exploiting $r_0$.}\label{fig:trader1}
\end{figure}

Trader $1$, a perfectly (and possibly illegally) informed speculator, will
take advantage of his knowledge by opening a position at time $t-1$ and
closing it at time $t+1$. It is important to be aware that if one places
an order at time $t$, the transaction takes place at price $p(t+1)$. Provided
that trader $0$ buys/sells $n_0$ shares irrespective of the price that
he obtains, the round-trip of trader $1$ yields a monetary gain of

\be
g_1=n_1[e^{p(t+2)}-e^{p(t)}]=n_1e^{p_0}[e^{I(n_0)}-e^{I(n_1)}].
\ee
and a relative gain of
\be
\hat g_1=\frac{n_1[e^{p(t+2)}-e^{p(t)}]}{n_1e^{p(t)}}=e^{I(n_0)-I(n_1)}-1,
\ee
\noindent where $p_0$ is the log-price before any trader considered here
makes a transaction. Since $I(n)$ generally increases with $n$, there is
an optimal $n_1^*$ number of shares  that maximises $g_1$.
The discussion so far is a simplification, in real-money instead of
log-money space,  of the one found in \citeasnoun{FarmerForce}. One
should note that the intervention of trader $1$ does not reduce price predictability (see Fig. \ref{fig:trader1}). Assuming that $|r_1|$ is equal to the market volatility, a straightforward computation shows that trader 1's actions increase the volatility measured in $\{t-1,t,t+1\}$ by a factor $9/8$, thereby reducing the signal-to-noise ratio around $t$. Therefore, in the
framework of constant price impact functions, an isolated arbitrage
opportunity never vanishes but becomes less and less exploitable because
of the fluctuations, thus the reduction, of signal-to-noise ratio caused
by the speculators.

It seems that trader $1$ cannot achieve a better gain than by holding
$n_1^*$ shares at time $t$. Since the actions of trader $1$ do not
modify in any way the arbitrage opportunity between $t-2$ and $t+2$, he
can inform a fully trusted friend, trader $2$, of the gain opportunity
on the condition that the latter opens his position before $t-1$ and
closes it after $t+1$ so as to avoid modifying the relative gain of
trader $1$.\footnote{If trader $2$ were not a good friend, trader $1$ could in principle ask trader $2$ to open his position after him and to close it after him, thus earning more. But relationships with real friends are supposed to egalitarian in this paper.} For instance, trader $2$ informs trader $1$ when he has
opened his position and trader $1$ tells trader $2$ when he has closed
his position. From the point of view of trader $2$, this is very
reasonable because the resulting action of trader $1$ is to leave the
arbitrage opportunity unchanged to $r_0$ since $p(t+1)-p(t-1)=r_0$.
Trader $2$ will consequently buy $n_2^*=n_1^*$ shares at time $t-2$ and
sell them at time $t+2$, earning the same return as trader $1$. This can
go on until trader $i$ has no other fully trusted friend. Note that the
advantage of trader $1$ is that he holds a position over a smaller time
interval, thereby increasing his return rate; in addition, since trader $2$
increases the opening price of trader $1$, which results into a
prefactor $e^{I(n_2)}$ in Eq.\ \req{eq:g1opt}, the absolute monetary
gain of trader $1$ actually {\em increases} provided that he has enough
capital to invest. Before explaining why this situation is paradoxical, it makes
sense to emphasise that the gains of traders $i>0$ are of course obtained
at the expense of trader $0$, and that the result the particular order
of the traders' actions is to create a bubble which peaks at time $t+1$.

The paradox is the following: if trader $1$ is alone, the best return
that can be extracted from his perfect knowledge is $\hat g_1(n_1^*)$
according to the above reasoning. When there are $N$ traders in the ring
of trust, the total return extracted is $N$ times the optimal gain of a
single trader. Now, assume that trader $1$ has two brokering accounts;
he can use each of his accounts, respecting the order in which
to open and close his positions, effectively earning the optimal return
on each of his accounts. The paradox is that his actions would be
completely equivalent to investing $n_1^*$ and then $n_1^*$ from the
same account. In particular, in the case of $I(n)=n$, this seems {\em a
priori} exactly similar to grouping the two transactions into $2n_1^*$,
but this results of course in a return smaller than the optimal return for
a doubled investment. Hence, in this framework, trader $1$ can earn as
much as  pleases  provided that he splits his investment into
sub-parts of $n_1^*$ shares whatever $I$ is, as long as it is constant.

Two criticisms can be raised. First, the ring of trust must be perfect
for this scheme to work, otherwise a Prisoner's dilemma arises, as it is
advantageous for trader $i+1$ to defect and close his position before
trader $i$. In that case, the expected return for each trader is of
order $1/N$, as in \citeasnoun{FarmerForce}.

But more importantly, one may expect that the above discussion relies
crucially on the fact that $r_0$ does not depend on the actual price, or
equivalently that trader $0$ wishes to buy or to sell a predetermined
number of shares. As we shall see in the second part of this paper, the paradox still exists even if trader
$0$ has a fixed budget $C$ (which is more likely to arise if trader $0$
intends to buy).

But this paradox seems too good to be present in real markets. As a consequence, one should rather consider its impossibility as an {\em a contrario}  consistency criterion for price impact functions. The final part of this paper explores the dynamics of price impact functions and the role of the spread in ensuring consistency, which is then tested in real markets.

\section{Finite capital}

When trader $0$ has a finite capital, the number of shares that he can
buy decreases when traders $1$, $2$, \dots increase the share price before his
transaction. Let us assume that trader $0$ has capital $C$ that buys $n_0$ shares at price $e^{p_0}$. The price he obtains is different from $e^{p_0}$ because of his impact: the real quantity of shares that $C$ can buy is therefore self-consistently determined by
\be
n_0^{(0)}=\frac{n_0}{e^{I(n_0^{(0)})}}.
\ee
We shall first focus on the case where the self impact is neglected, or equivalently where trader $0$ has a restricted budget with flexible constraint since it is leads to simpler mathematical expressions.

When  only trader $1$ is active before trader $0$, $n_0$ becomes $n_0^{(1)}=n_0/e^{I(n_1)}$, thus the gain of trader $1$ is
\be\label{eq:g1}
g_1/e^{p_0}=n_1[e^{I(n_0/e^{I(n_1)})}-e^{I(n_1)}].
\ee

It is easy to convince oneself that there is always an $n_1^*\ge1$ provided that $n_0$ is large enough: solving $\partial g_1/\partial n_1=0$ and setting $n_1=1$ gives the corresponding $n_0$. Using the shortcut $\rho=e^{I(1)}$,
\be
e^{I(n_0/\rho)}\cro{1-\frac{n_0}{\rho^2}I'(1)I'(n0/\rho)}-\rho[1+I'(1)]=0
\ee
The existence of a solution to this equation is intuitive: when $n_0=0$, the left hand side is negative; since the second l.h.s term does not depend on $n_0$, one is left to show that the first l.h.s term grows fast enough with $n_0$ so that its parenthesis does vanish. Since $I'(x)$ is assumed to be small in financial markets, the second l.h.s term is close to 1. Now, assuming that $I(x)$ grows at least as fast as a $\log x$, the first term always gives a solution. The case $I(x)\propto\log x$ is dealt with explicitly in the next subsection. At the other end of the spectrum of usual price impact functions, the linear case $I(x)=\lambda x$ cannot be solved exactly, but assuming that $\lambda\ll1$ and perusing $e^{I(x)}\simeq 1+\lambda x$, one finds $n_0\simeq \frac{1+\lambda}{2\lambda^2}$, which deviates by about 3\% from the numerical solution for $\lambda=0.01$.

Trader $1$ must now be careful when communicating the existence of the arbitrage to trader $2$, since the latter decreases the price return caused by trader $0$. Indeed, assuming that trader $1$ invests $n_1^*$ whatever trader $2$ does, his gain is given by
\be\label{eq:g12}
g_{1,2}(n_1^*,n_2)/e^{p_0}=n_1^*[e^{I(n_2)+I(n_0/e^{I(n_1^*)+I(n_2)})}-e^{I(n_2)+I(n_1^*)}].
\ee
He would therefore lose $\Delta g_1(n_2)=g_1^*-g_{1,2}(n_1^*,n_2)$, which must be compensated for by trader $2$. Without compensation payment, the gain of the latter is
\be\label{eq:g2}
g_{2}/e^{p_0}=n_2[e^{I(n_0/e^{I(n_1^*)+I(n_2)})}-e^{I(n_2)}].
\ee
The case where trader $0$ has a strict budget constraint is obtained by replacing $n_0$ by $n_0^{(0)}$ in Eqs \req{eq:g1}, \req{eq:g12} and \req{eq:g2}.

The paradox exists if trader $2$ has a positive total gain, i.e.,  $G_2=g_2-\Delta g_1>0$. In order to investigate when this is the case, one must resort to particular examples of price impact functions.

Empirical research showed that $I$ is a non-linear, concave function \cite{Hasbrouck91,KempfKorn99,FarmerImpact,BouchaudLimit2,RosenowImpact}. Although there is no consensus on its shape, logarithms and power- laws are possible candidates. From a mathematical point of view, logarithmic functions allow for explicit computations, while power-laws must be investigated numerically.

\subsection{Log price impact functions}

The generic impact function that will be studied is $I(|x|)=\sign(x) \gamma \log |\lambda x|$; when $\lambda>1$, a transaction of one share results in a price change; the number of shares will be rescaled so as to
contain $\lambda$, which shorten the mathematical expressions. The parameter $\gamma\ll1$ is related to the liquidity and brings down the impact to reasonable levels: if $x=100$ and $\gamma=0.01$, the price is increased by about $5$\%.

For the sake of comparison, we first address the case of infinite capital: the optimal number of shares to invest by traders $1$ and $2$ is
\be
n_1^*=\frac{n_0}{(\gamma+1)^{1/\gamma}},
\ee
\noindent
which simplifies to $n_0/2$ if $\gamma=1$, and tends to $n_0/e^{1+\gamma/2}$ in the limit of small $\gamma$. The optimal monetary gain is given by
\be\label{eq:g1opt}
g_1^*=e^{p_0}n_0^{\gamma+1}\frac{\gamma }{(\gamma+1)^{1+1/\gamma}}.
\ee

In the case of finite capital, the optimal number of shares to invest for trader  $1$ is
\be\label{eq:n_1starC}
n_1^*=\cro{n_0^\gamma(1-\gamma)}^{\frac{1}{\gamma(\gamma+1)}};
\ee

The optimal gain is given by

\be
g_1^*=e^{p_0}n_0\frac{\gamma}{(1-\gamma)^{1-1/\gamma}}=C\frac{\gamma}{(1-\gamma)^{1-1/\gamma}},
\ee

\noindent which is linear in $n_0$, in contrast to its super-linearity
in Eq \req{eq:g1opt}: the limited budget of trader $0$ helps to remove
predictability.

The gain of trader $1$ in the presence of trader $2$ is

\be
g_{1,2}(n_1^*,n_2)=e^{p_0}n_0(1-\gamma)^{1/\gamma}\pr{\frac{1}{(1-\gamma)n_2^{\gamma^2}}-1};
\ee
he therefore would lose
\be
\Delta g_1(n_2)=g_1^*-g_{1,2}(n_1^*,n_2)=e^{p_0}n_0(1-\gamma)^{1/\gamma-1}\cro{1-\frac{1}{n_2^{\gamma^2}}}.
\ee

\begin{figure}
\centerline{\includegraphics[width=0.4\textwidth]{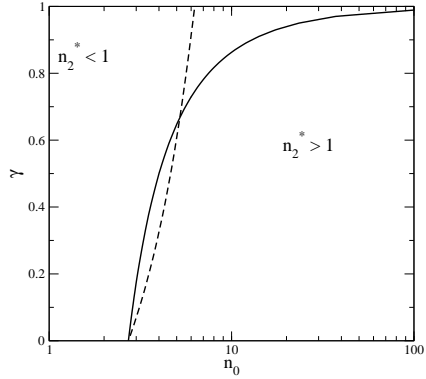}}
\caption{Region where the paradox exists ($n_2^*>1$)  in
parameter space ($n0$, $\gamma$). Logarithmic impact, finite capital with flexible constraint (continuous line) and strict constraint (dashed line).}

\label{fig:g2tilde}
\end{figure}

Trader $2$ optimises

\be
\frac{G_2}{e^{p_0}}=[g_2-\Delta g_1]e^{-p_0}=\cro{\frac{n_0}{1-\gamma}}^{\frac{\gamma}{(\gamma+1)}}n_2^{1-\gamma^2}-n_2^{1+\gamma}-\Delta g_1(n_2)
\ee

\noindent with respect to $n_2$. The paradox survives in the regions of
the parameter space such that $G_2>0^*$ and $n_2^*> 1$. Numerical investigations show that $G_2^*>0$ is always true. From Fig.\ \ref{fig:g2tilde} one sees that the paradox exists provided that scaled $n_0$ is large enough ($\lambda n_0>e^{1+\gamma/2}$ for $\gamma\ll1$) and $\gamma$ small enough, which is compatible with all realistic values. Note that $g_1^*$ and $G_2^*$ are increasing functions of $\gamma$, since this parameter tunes the price return caused by $n_0^{(1)}$ and $n_0^{(2)}$.

If trader $0$ has a strict budget, the number of shares he can afford is determined self-consistently by $n_0^{(0)}=C/e^{p_0+\gamma\log n_0^{(0)}}$  without the intervention of the trader $1$, that is, $n_0^{(0)}=n_0^{1/(\gamma+1)}$; after trader $1$ has opened his position, trader $0$ can only afford
 \be\label{eq:n01}
n_0^{(1)}=n_0^{(0)}/n_1^{\gamma/(\gamma+1)}=\cro{\frac{n_0}{n_1^\gamma}}^{\frac{1}{\gamma+1}}
\ee
 shares. Similarly,
\be
n_1^*=n_0^{\frac{1}{2\gamma+1}}\cro{\frac{1-\gamma^2/(\gamma+1)}{\gamma+1}}^{\frac{\gamma+1}{\gamma(2\gamma+1)}},
\ee

\noindent hence the number of shares to invest is lowered further in
comparison with Eq \req{eq:n_1starC}. All the expressions of the
respective gains of trader $1$ and $2$ do not simplify to neat and short
expressions and no analytical solution to the maximisation of
$G_2$ can be found.  Numerical maximisation yields the  dashed boundary line in Fig.\ \ref{fig:g2tilde}. The strict constraint increases the minimum $n_0$ needed for making the paradox possible when $\gamma$ is small enough. At $\gamma\simeq 0.66$, the boundary lines cross; this may be due to the fact that trader $2$ must pay a smaller compensation to trader $1$ when $\gamma$ increases.

\subsection{Power-law impact functions}

Several papers suggest a power-law price impact function $I(x)=\sign x |x/\lambda|^\beta$ with $\lambda\ge1$ and $\beta\in[0.3,0.8]$  \cite{StanleyImpact,RosenowImpact,FarmerImpact}; for the sake of simplicity (and without loss of generality) the volumes $x$ will be rescaled by $\lambda$ from now on; this means that the consistency conditions are now $G_2>0$ and $n_2>0$. With this choice of functions, it is not possible to maximise explicitly $g_1$ and  $G_2$.

\begin{figure}

 \centerline{\includegraphics[width=0.4\textwidth]{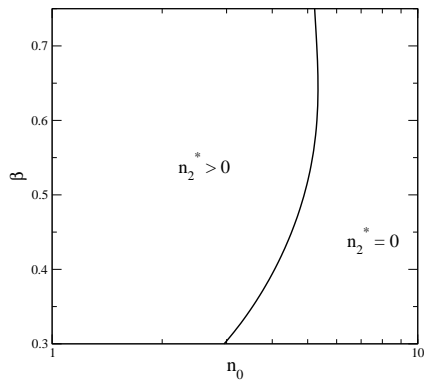}}
\caption{Region where the optimal gain of trader $2$ is positive in the (scaled) $n_0$, $\beta$ space. Power-law impact function and finite capital.}\label{fig:g1g2-power}
\end{figure}

Performing all the maximisations of the finite capital case numerically, one finds that the picture of the $\log$ impact function is still valid in this case (see Fig.\ \ref{fig:g1g2-power}): realistic values of (scaled) $n_0$ are small, hence they always fall into the region of inconsistency, in particular for $I(x)\propto x$. Interestingly, this is the most common choice in the literature on agent-based models, option pricing, etc. It was derived analytically by \citeasnoun{Kyle85} under the assumption of linearity between private information and insiders' order flow. More recently, \citeasnoun{FarmerForce} uses it as simple example of a function that prevents making trading profits from a round trip. And certainly, it may seem the least arbitrary, since it does not seem to impose any particular choice of $\beta$.  Hence constant $I(x)\propto x$  is to be banished. But power-laws are inconsistent for all possible real-market values of $\beta$; once again, this shows that constant price impact functions are to be avoided.

\section{Feedback}

A possible way out from inconsistent price impact functions is to take into account the dynamics of the order book, particularly the reaction of the order book to a sequence of market orders of the same kind. Generically, the impact of a second market order of the same kind and size is smaller than that of the first one, and similarly for a third one. For our purpose, we shall assume that a second market order of the same size and type has an impact described by $I_1(n)=\kappa_1 I(n)$ with $0<\kappa_1<1$, the third $I_{2}(n)=\kappa_2\kappa_1 I(n)$, etc \cite{BouchaudLimit5,FarmerMolasses}. To this contraction of market impact on one side also corresponds an increase of market impact on the other side for the next market order of opposite type \cite{BouchaudLimit5}; therefore, we shall assume that the impact function on the other side is divided by $\theta_1$ after the first market order, by $\theta_2\theta_1$ after the second, etc. As shown in the next section, $\kappa_1\simeq\kappa_2$ is a very good approximation when $\kappa_1$ and $\kappa_2$ are averaged over all the stocks, hence we shall assume $\kappa_1=\kappa_2=\kappa$; for the same reason, we assume that $\theta_1=\theta_2=\theta$. The hypothesis that the contractions of the price impact functions correspond to multiplying $I$ for each new market order makes this process Markovian. While it is now well established that order books and order flows are not Markovian but display long memory \cite{FarmerMolasses,BouchaudImpact2}\footnote{A discussion about why Markovian models such as \citeasnoun{Madhavan} are inherently inadequate to describe real order book dynamics is found in \cite{BouchaudLimit5}}, this may seem a step back. However since we  only consider a few time steps, the long memory of the order book is of little importance for the present discussion.

When writing the gains of trader $1$ and $2$, one must be very careful with the order of the transactions. Starting with the case of infinite capital of trader $0$, the gain of trader $1$ is
\be
g_1=n_1[e^{I(n_1)+\kappa I(n_0)-I(n_1)/\theta^2}-e^{I(n_1)}],
\ee
which makes it clear that there is still a non-zero optimal number of shares $n_1^*$ to invest. Similarly, the gain $g_{1,2}$ is now
\be
g_{1,2}=n_1^*[e^{I(n_2)+\kappa I(n_1^*)+{\kappa^2}I(n_0)-I(n_1^*)/\theta^3}-e^{I(n_2)+\kappa I(n_1^*)}],
\ee
while that of trader 2, without his payment to trader 1 is
\be
g_{2}=n_2(e^{I(n_2)+\kappa I(n_1^*)+{\kappa^2}I(n_0)-I(n_1^*)/\theta^3-\kappa I(n_2)/\theta^2}-e^{I(n_2)}).
\ee

\subsection{Feedback on market order side only}

In order to investigate whether the feedback on the side on which the first sequential market orders are placed is enough to make price impact consistent, one sets $\theta=1$. In the case of $\log$ price impact functions, the optimal number of shares and gain of trader $1$ are
\be
n_1^*=\frac{n_0^\kappa}{(\gamma+1)^{1/\gamma}}
\ee
and
\be\label{eq:g1optdyn}
g_1^*=e^{p_0}n_0^{\kappa(\gamma+1)}\frac{\gamma }{(\gamma+1)^{1+1/\gamma}}.
\ee
These two equations already show  that the reaction of the limit order book reduces the gain opportunity of player $1$. Adding trader $2$ will reduce further the impact of trader $0$, hence the gain of trader $1$, and, as before, trader $2$ should pay for it. In this case, the reduction of gain of trader $1$ is
\bea
\frac{\Delta g_1}{e^{p_0}}&=&[g_1^*-g_1(n_1^*,n_2)]e^{-p_0}=n_0^{\kappa(\gamma+1)}\frac{\gamma}{(\gamma+1)^{1+1/\gamma}}\nonumber\\&&-n_2^\gamma\frac{n_0^{\gamma(1+\kappa\gamma)}}{(\gamma+1)^{1/\gamma+\kappa}}[n_0^{-\gamma\kappa(1-\kappa)}(\gamma+1)-1],
\eea
while the gain that trader $2$ optimises is
\be
\frac{G_2}{e^{p_0}}=n_2^{\gamma+1}\cro{\frac{1}{n_2^{\kappa\gamma}}\frac{n_0^{\kappa\gamma(2\kappa-1)}}{(1+\gamma)^{\kappa-1}}-1}-\frac{\Delta  g_1}{e^{p_0}}.
\ee

\begin{figure}
 \centerline{\includegraphics[width=0.4\textwidth]{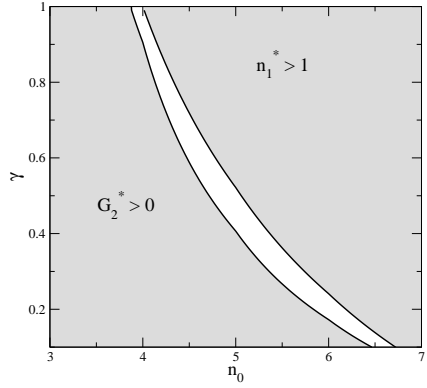}}
\caption{Regions in which the optimal gain of trader $2$ is positive and in which the number of shares invested by trader $1$ is larger than $1$. Logarithmic impact, infinite capital, and feedback on market orders side only ($\kappa=0.5$).}\label{fig:G2n1}
\end{figure}

Trader $1$'s impact functions are $\kappa I$ when he opens his position and $I$ when he closes it, which is an additional cause of loss for trader $1$, which must be also compensated for by trader $2$. Fortunately for the latter, his impact functions are $I$ when opening and $\kappa I$ when closing his position. Therefore, provided that $\kappa$ is large enough so as not to make the impact of the trader 0 $\kappa^2I(n_0)$ too small, trader $2$ can earn more than trader $1$ in some circumstances.

As above, impact functions are inconsistent when $G_2^*>0$, $n_1^*>1$ and $n_2^*>1$ for $\log$ impact functions. It turns out that the regions in which $G_2^*>0$ while $n_1^*>1$ are disjoint if $\kappa<\kappa_c\simeq 0.5$ for $\log$ impact functions (Fig. \ref{fig:G2n1}).

\begin{figure}
 \centerline{\includegraphics[width=0.4\textwidth]{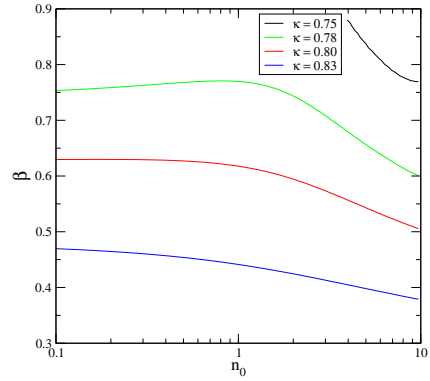}}
\caption{Regions of parameter space where the optimal gain of trader $2$ ($G_2^*$) is positive (above the respective line). Power-law impact functions, feedback with infinite capital. From top to bottom $\kappa=0.75$, $0.78$, $0.8$ and $0.83$. }\label{fig:power-fb-G2gt0}
\end{figure}

The case of power-law impact functions shows a similar transition: since $n_0$ is rescaled, the realistic region to consider is limited to $n_0\ll 1$, for which the region $G_2^*>0$ vanishes when $\kappa<\kappa_c\simeq 0.78$ (Fig.\ \ref{fig:power-fb-G2gt0}).

\begin{figure}
 \centerline{\includegraphics[width=0.4\textwidth]{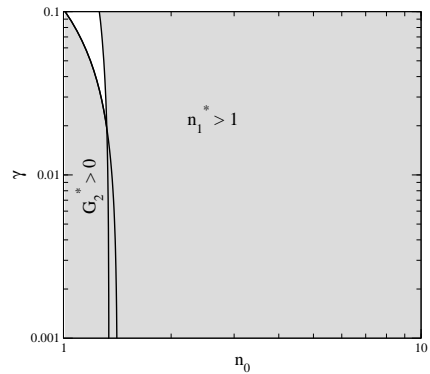}}
\caption{Regions in which the optimal gain of trader $2$ is positive and in which the number of shares invested by trader $1$ is larger than $1$. Logarithmic impact, infinite capital, and feedback on both sides  ($\kappa=0.83$)}\label{fig:log-fb2}
\end{figure}

\begin{figure}
 \centerline{\includegraphics[width=0.4\textwidth]{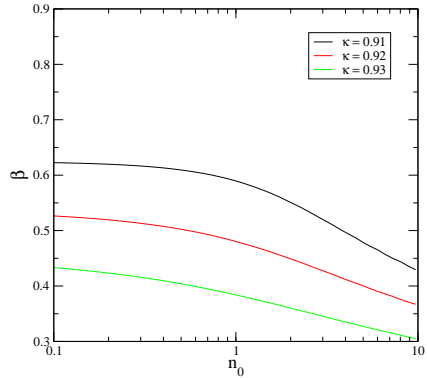}}
\caption{Regions in which the optimal gain of trader $2$ is positive (above the respective line). Power-law impact, infinite capital, and feedback on both sides; from top to bottom: $\kappa=0.91$ (black line), $\kappa=0.92$ (red line), $\kappa=0.93$ (green line).}\label{fig:power-fb2}
\end{figure}

\subsection{Feedback on both sides}

We shall assume that $\theta_1=\theta_2=\kappa$ in order to be able to use market data. Generalising the above calculations is straightforward. As an example, for $\log$ price impact functions,
\be
n_1^*=n_0^{\kappa^3}\cro{\frac{1-\gamma(1/\kappa^2-1)}{1+\gamma}}^{\kappa^2/\gamma}.
\ee
Once again, more feedback results in a smaller exponent of $n_0$. As a consequence, Fig.\ \ref{fig:log-fb2} shows that the critical value of $\kappa$ is slightly larger than $0.83$ for $\log$ price impact functions all  $\gamma$: indeed only a small area of inconsistent impact functions, corresponding to $n_0\simeq 1.5$, still exists but cannot be reached since both $n_0=1$ and $2$ are outside of the inconsistent region. The effect of feedback on power-law price impact functions is stronger (see Fig.\ \ref{fig:power-fb2}): when $\kappa<0.92$, (scaled), only power-laws with exponent larger than about $0.5$ (this includes linear and super-linear impact functions) are inconsistent for reasonable values of $n_0$. Therefore, even double feedback does not guarantee consistency

\section{Empirical data}

\begin{figure}
 \centerline{\includegraphics[width=0.4\textwidth]{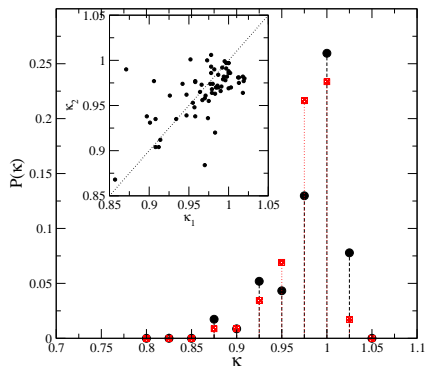}}
\caption{Empirical distribution of $\kappa_1$ (black circles) and $\kappa_2$ (red squares) of 68 stocks of Paris Stock Exchange in 2002. Inset: $\kappa_2$ vs $\kappa_1$ for all the stocks}\label{fig:Pkappa}
\end{figure}

The values of $\kappa_1$ and $\kappa_2$ can be measured in real markets.
The response function $R(\delta t,V)=\avg{(p(t+\delta t)-p(t))\eps(t)|V(t)=V}$ is the average mid-price change after $\delta t$ trades, conditional on the sign of the trade $\eps(t)$ and on volume $V$; similarly, one defines the response function conditional on two trades of the same sign $R^+(\delta t,V)=\avg{(p(t+\delta t)-p(t))\eps(t)|V(t)=V,\eps(t)=\eps(t-1)}$, and $R^{++}(\delta t,V)=\avg{(p(t+\delta t)-p(t))\eps(t)|V(t)=V,\eps(t)=\eps(t-1)=\eps(t-2)}$. A key finding of \citeasnoun{BouchaudLimit3} and \citeasnoun{BouchaudLimit4} is that $R$ factorises into $R(\delta t)F (V)$. Thus we will be interested in $R(\delta t)$, $R^+(\delta t)$ and $R^{++}(\delta t)$.

Using measures kindly provided by J.-Ph. Bouchaud and J. Kockelkoren taken on one year of data about all the order books of Paris Stock Exchange, one finds that the estimate $\hat \kappa$ of $\kappa$, defined as  $\hat \kappa=\avg{R^+(1)}/\avg{R(1)}\in[0.86,1.02]$, that the average over all stocks $\ovl{\hat\kappa_1}=0.97\pm0.04$ and $\hat \kappa_2=\avg{R^{++}(1)}/\avg{R^+(1)}\in[0.87,1.01]$, while $\ovl{\hat\kappa_2}=0.97\pm0.03$. (see Fig.\ \ref{fig:Pkappa}); for a given stock, there is some correlation between $\kappa_1$ and $\kappa_2$ (see inset of Fig.\ \ref{fig:Pkappa}); the data presented here does not contain error bars for the measures of $R$, $R^+$ and $R^{++}$. The approximation $\kappa_1\simeq \kappa_2$ is reasonable, and we shall from now on call $\kappa=(\kappa_1+\kappa_2)/2$ and replace $\kappa_1$ and $\kappa_2$ by $\kappa$ everywhere. It should be noted that since one considers the mid-price reaction to a market order, in principle, the measured values of $R$, $R^+$, and $R^{++}$ include the spread. However, because $\kappa$ is the ratio between two of them, this effect cancels out.\footnote{Mathematically, if $a(t)$ and $b(t)$ denote the ask and bid prices, $m(t)=(a+b)(t)/2$. Hence, focusing on a buy market order of size $V_t$ at time $t$ and assuming that it leaves $b(t)$ unchanged, $m(t+1)-m(t)=(a'-a)(t+1)/2$. If $a'(t)=a(t)+Q(t)F(V(t))$, $m(t+1)-m(t)=Q(t)F(V(t))/2$, where $Q$ is the function that describes the impact of this precise trade and contains all the memory and state of the ask order book at time $t$; now $\kappa_1$ is nothing else than $\avg{\frac{Q(t)}{Q(t-1)}}|_{\eps(t)=\eps(t-1),V(t)=V(t-1)}$; the fact that one takes the ratio of $R$ and $R^+$ removes the factor 1/2 (a similar argument holds for $\kappa_1$); this ensures that the spread is not taken into account in the measurement of $\kappa_1$ and $\kappa_2$.}

In other words there is some variations between the stocks, some of them being less sensitive to successive market orders of the same kind. The values of estimated $\kappa$ start at $0.86$. Therefore, even feedback on both book sides does not yield consistent $\log$ impact functions, but do if the impact functions are power-laws with small exponents. One concludes that the feedback of the order book is not enough to make price impact functions consistent.

\section{Spread without feedback, infinite capital}

\begin{figure}
 \centerline{\includegraphics[width=0.4\textwidth]{avgV_vs_n0min.eps}}
\caption{Average daily volume versus $n_{0,min}$ (same data set)}\label{fig:avgV_vs_n0min}
\end{figure}

\begin{figure}
 \centerline{\includegraphics[width=0.4\textwidth]{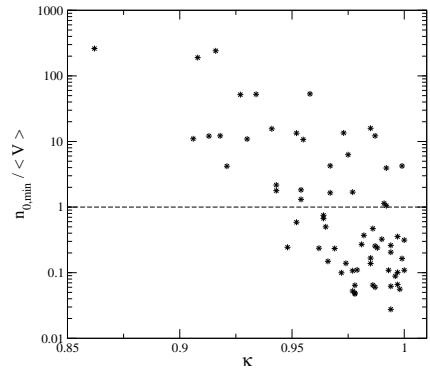}}
\caption{Fraction of daily volume needed to inject an exploitable arbitrage versus $\kappa$ (same data set) }\label{fig:n0mindivavgV_vs_kappa}
\end{figure}

The above discussion neglects the bid-ask spread $s$. It is however of great importance in practice, as the impact of one trade is on average of the same order of magnitude as the spread \cite{BouchaudLimit5}. This means that $n_0$ must be large enough in order to make the knowledge of trader $1$ valuable. Assuming that the spread is equal to $\avg{s}$ at all times, Eq.\ \req{eq:g1} generalises to
\be
\frac{g_1}{e^{p_0}}=n_1\cro{e^{I(n_0)-\avg{s}}-e^{I(n_1)}}.
\ee
Thus, replacing $n_0$ by $n_0'=I^{-1}[I(n_0)+\avg{s}]$ leads to the same equations as before.   For example, in the case of $log$ impact functions, $n_0'=n_0e^{-\frac{\avg{s}}{\gamma}}$ and the optimal number of shares that trader $1$ should invest is
\be
n_{1,s}^*=\frac{n_0 e^{-\avg{s}/\gamma}}{(1+\gamma)^{1/\gamma}}
\ee
and his optimal gain is given multiplied by $e^{-\frac{1+\gamma}{\gamma}\avg{s}}\simeq e^{-\frac{\avg{s}}{\gamma}}$.
Since $\avg{s}/\gamma\sim10$ in practice, the minimal amount of shares needed to create an arbitrage, denoted by $n_{0,\min}$ is increased (and his gain divided) about 20,000 folds by the spread. Straightforward calculations show that replacing  $n_0$ by $n_0'$ holds for all the equations investigated in previous subsections.

The respective values of $\gamma$ and $\avg{s}$ are not independent, and can be measured in real markets for a given stock. In the language of \citeasnoun{BouchaudLimit3}, $\gamma=\avg{\log(n)}/R(1)$ where $\avg{\log (n)}$ is the average of the logarithm of transaction size and $R(1)$ is the response function after one time step. Using $\gamma$ and $\avg{s}$ measured in Paris Stock Exchange one finds that $n_{0,\min}\in[1.2 10^4,2.0 10^8]$, with median of  $1.2 10^5$ (Fig.\ \ref{fig:avgV_vs_n0min}), which is not irrealistic for very liquid stocks. Indeed, the fraction of $n_{0,\min}$ with respect to  the average daily volume of each stock ranges from $\simeq2\%$ to more than $100\%$, with a median of about $42\%$. Therefore, for most of the stocks, trader $0$ needs to trade less than two fifths of the daily average volume in one transaction in order to be leave an exploitable arbitrage; for 12 stocks ($18\%$), trading less than $10\%$ of the average daily volume suffices. It is unlikely that a single trade is larger than the average daily volume, hence, 29 stocks ($43\%$) do not allow {\em on average} a single large trade to be exploitable by a simple round-trip. Interestingly, stocks with average daily volumes smaller than about $10^5$ are all consistent from that point of view (Fig.\ \ref{fig:avgV_vs_n0min}). In addition, the stocks for which $n_{0,\min}<\avg{ V}$ all have a $\kappa>0.945$.

Therefore, the role of the spread is to increase considerably the minimum size of the trade, which in some cases remain within reasonable bounds. Therefore, the spread must be taken into account, but does not yield systematically consistent impact functions for some stocks with a high enough daily volume.

\section{Spread and feedback, infinite capital}

The question is whether the feedback and the spread make impact function systematically consistent. The stocks that are the most likely to become consistent are those whose $n_{0,\min}/\avg{ V}<1$ is large while having a strong feedback. According to  Fig.\ \ref{fig:n0mindivavgV_vs_kappa}, these properties are compatible.

Using for each stock $\avg{s}$, $\kappa$, and $\gamma$ from the data, we find that three additional stocks (ALS, IFG, SCO) are made consistent by feedback on trader $0$'s market order side alone: the feedback limited to one side of the order book, even when the spread is taken into account, is insufficient. However, adding finally the feedback on both book sides makes consistent {\em all} the stocks, even in the case of infinite capital. Therefore, both the spread and the feedback on both sides of the order book are crucial ingredients of consistency at the smallest time scale.

\section{Discussion}

 The paradox proposed in this paper provides an {\em a contrario} simple and necessary condition of consistency for price impact functions. This condition is verified for {\em all} the stocks of Paris Stock Exchange in 2002.  Indeed, the main finding of this paper is that financial markets ensure consistent market price impact functions on average at the most microscopic dynamical level by two essential ingredients: the spread and the dynamics of the order book on both sides of the order book.

The fact that non-constant price impact functions with feedback are needed to achieve consistency at the smallest time scale for more than half of the stocks means that financial literature that assumes constant price impact functions must be critically reviewed in this light. This paper is not the first to suggest it (see e.g \cite{BouchaudLimit3,BouchaudLimit4,FarmerImpact,FarmerMolasses}), but advocates it from logical evidence: for instance \citeasnoun{HubermanStanzl} shows that  if the price impact is permanent, only linear impact functions prevent quasi-arbitrage, but recognises the paradox that real-market price impact functions are not linear, therefore suspects the need for time-dependent impact functions. The approach of the present paper remains Markovian, which is justified when restricting one's attention to the smallest time scale and a few transactions only. However, in general, the long memory of the side of limit order markets must be compensated in a subtle way by the price impact functions (see e.g. \citeasnoun{BouchaudLimit4} and \citeasnoun{FarmerMolasses}). This raises the crucial issue of how to incorporate in a tractable way non-constant impacts in microstructure and agent-based modelling.
\bigskip

I am indebted to Jean-Philippe Bouchaud, Julien Kockelkoren and Michele Vettorazzo from CFM for their hospitality, measurements,  and comments. It is a pleasure to acknowledge discussions and comments from Doyne Farmer, Sam Howison, Matteo Marsili, Sorin Solomon and David Br\'ee, and useful criticisms by an anonymous referee.
\bibliographystyle{kluwer}
\bibliography{biblio}

\end{document}